\documentclass[reprint,amsmath,amssymb,aps,pra,longbibliography]{revtex4-1}

\usepackage{graphicx}
\usepackage{dcolumn}
\usepackage{bm}
\usepackage[colorlinks=true, allcolors=black]{hyperref}
\usepackage{breakurl}  
\usepackage[mathscr]{euscript}

\DeclareMathSymbol{\mlq}{\mathord}{operators}{``}

\newcommand{\I}{\mathrm{i}}

\begin{document}

\title{Strong-field photoelectron holography beyond the electric dipole approximation:\\
	a semiclassical analysis}

\author{Simon Brennecke}
\affiliation{%
	Institut f\"ur Theoretische Physik, Leibniz Universit\"at Hannover, Appelstra{\ss}e 2, 30167 Hannover, Germany
}
\author{Manfred Lein}%
 \email{lein@itp.uni-hannover.de}
\affiliation{%
Institut f\"ur Theoretische Physik, Leibniz Universit\"at Hannover, Appelstra{\ss}e 2, 30167 Hannover, Germany
}%

\date{\today}

\begin{abstract}
Strong-field photoelectron holography denotes the 
interference of various electron paths in laser-induced ionization of
atoms, leading to interference patterns in the final momentum distribution,
typically characterized by a strong signal of momenta pointing
along the laser polarization axis and interference fringes roughly
parallel to the polarization axis.
For a quantitative trajectory-based description of holography
beyond the electric dipole approximation and in the presence of the
Coulomb potential, we develop a semiclassical model in which
the initial conditions of outgoing electrons
are set according to the beyond-dipole strong-field
approximation for the tunnel ionization step. The phases associated
with electron trajectories are evaluated following the prescription
for semiclassical propagators. The validity of the method is confirmed
by comparison to the numerical solution of the time-dependent
Schr\"odinger equation in two spatial dimensions. The
semiclassical model reproduces
correctly the backward and forward shifts of the photoelectron momenta
along the laser propagation axis that arise from beyond-dipole dynamics.
The position of the central holographic interference fringe can be
estimated already from a simplified Coulomb-free
interference model that provides closed-form expressions for the beyond-dipole shifts.
In three dimensions, Coulomb focusing causes a breakdown of the semiclassical model for
final momenta with directions close to the polarization axis. We implement
a beyond-dipole regularization procedure based on the concept of
glory scattering, which was recently used to describe Coulomb focusing
in the dipole approximation. While the position of the central maximum and higher-order fringes in three dimensions can already be obtained approximately by simpler semiclassical modeling, this glory model is able to predict the shape of the distribution in the close vicinity of the central maximum. Our results show that the violation of the
dipole approximation in holography should be observable with mid-infrared fields,
for which the forward/backward shifts can be comparable with the
fringe spacing.
\end{abstract}

\pacs{Valid PACS appear here}

\maketitle

\section{Introduction}
The idea of conventional holography \cite{gabor1992} can be transferred to strong-field ionization to selfprobe atomic and molecular structure and dynamics on a subfemtosecond time scale \cite{huismans2011}. In strong-field photoelectron holography (SFPH), the electron wave packet after tunnel ionization is steered by the time-dependent laser field. The wave packet contains one part with large initial transversal velocities, forming a reference wave while another part with small transversal velocities is driven back to the ion where it scatters and forms the signal wave. The interference of the signal and reference waves creates a hologram that encodes spatial and temporal information about the ion as well as the recolliding electrons with attosecond resolution. Even though in the first experimental realizations, free-electron lasers were used \cite{huismans2011,huismans2012}, SFPH can be realized in table-top experiments \cite{marchenko2011,hickstein2012}. As an ultrafast imaging method it has been used to successfully probe the continuum electron phase of molecular ionization \cite{meckel2014} as well as the nuclear motion and also electronic valence-shell dynamics in molecules \cite{haertelt2016,walt2017}.

For the decoding of the photoelectron hologram and hence extraction of information on the investigated system, an appropriate modeling of the strong-field driven quantum dynamics is essential. The simplest possible approaches are the simple man's models \cite{bian2011,bian2012} that divide the full process into a sequence of several steps. The reference electrons can be described in a two-step model consisting of (i) laser-induced ionization and (ii) potential-free acceleration of the electron as a classical particle in the laser field \cite{gallagher1988,corkum1989}. This description predicts the well-known electron-energy cut-off at $2U_p$ for ``direct'' above-threshold ionization (ATI) \cite{agostini1979}, where $U_p=E_0/(4\omega^2)$ is known as the ponderomotive potential for a linearly polarized laser field with amplitude $E_0$ and photon energy $\omega$. (We use atomic units unless stated otherwise.) For large initial transversal velocities with respect to the laser polarization, the electron makes a wide turn around the ion. It interacts only weakly with the core and forms the reference wave. In contrast, for small transversal velocities the electron can be driven back to the close vicinity of the ion during its acceleration and it may scatter off the potential before it is subsequently accelerated in the field for a second time. The returning electron may also recombine with the parent ion and emit high-order harmonics \cite{mcpherson1987,ferray1988} or it can release another electron from the atomic ion, leading to nonsequential double ionization \cite{fittinghoff1992}. 
Due to this recollision step we speak of the three-step model \cite{kulander1993,corkum1993}. In SFPH, however, we focus on the case that the returning electron undergoes elastic scattering. ``Backscattered'' electrons form in photoelectron energy distributions a high-energy plateau reaching up to the $10U_p$ cutoff \cite{paulus1994,paulus1994a}. On the other hand, ``forward scattered'' electrons have much higher emission probability such that they typically form the signal wave packet in SFPH.  

For photoelectrons with low transversal final momenta this simple picture breaks down in long-range Coulomb-like potentials of positively charged ions. Here, in three dimensions an infinite number of classical trajectories, which are launched with nonzero initial velocities distributed on a specific circle in the plane of transversal momenta is mapped to a tiny region in the space of final momenta. Since the electrons are focused in the momentum space along the polarization axis the effect is called Coulomb focusing \cite{brabec1996,comtois2005}. For a quantitative description, the unphysical classical caustic has to be regularized by taking the quantum nature of the focusing process in terms of glory rescattering into account \cite{dewittmorette1997,xia2018}. The Coulomb potential manifests itself also in the rich low-energy structures near the ionization threshold in midinfrared laser fields \cite{blaga2009}. These low-energy structures have been traced back to multiple forward scatterings by the Coulomb potential during the oscillatory motion in the laser field \cite{liu2010,kästner2012}.

Photoelectron momentum distributions (PMDs) obtained by the numerical solution of the three-dimensional time-dependent Schr\"odinger equation (TDSE) agree well with experimental data. For example, the spacings between different side lobes (fringes) of the holographic pattern are reproduced \cite{huismans2011,huismans2012}. However, such calculations are only possible for small systems with high symmetries in a restricted region of laser parameters. A simplified quantum-mechanical theory known as strong-field approximation \cite{keldysh1964,faisal1973,reiss1980} can be obtained by assuming that once the electron has been released, its motion is fully governed by the laser field. This method describes successfully ``direct'' ionization. However, even a perturbative inclusion of the ionic potential by means of a Born series \cite{becker1994,lohr1997} cannot model correctly SFPH in realistic long-range potentials. Employing classical trajectories prescribed by Newton's equation for the acceleration step of the two-step model, the classical trajectory Monte Carlo (CTMC) simulations can be used to study qualitatively Coulomb focusing as well as low-energy structures \cite{kästner2012}. Furthermore, semiclassical models such as quantum trajectory Monte Carlo (QTMC) simulations \cite{li2014}, the coulomb-corrected strong-field approximation (CCSFA) \cite{yan2010,popruzhenko2008,popruzhenko2008a}, the semiclassical two-step model (SCTS) \cite{shvetsov2016} or the Coulomb quantum orbit strong-field approximation (CQSFA) \cite{lai2015} offer the possibility to describe interference effects by associating to each classical trajectory a phase determined by the classical action. However, the CCSFA and QTMC models include the Coulomb potential only in first-order perturbation theory in the semiclassical phase such that the positions and spacings of interference structures are only qualitatively reproduced \cite{shvetsov2016}. The SCTS and CQSFA approaches are based on a semiclassical approximation of the time-dependent propagator formulated as Feynman path integral (see Ref. \cite{kleinert2009} for a textbook treatment) such that their description of the second step includes the Coulomb potential in a non-perturbative manner.

The description of the introduced processes is mostly carried out in electric dipole approximation where it is assumed that the incident electric field is spatially homogeneous over the field-target interaction region and furthermore the magnetic field is completely neglected. The relevant parameter that characterizes the applicability of this approximation is the relativistic parameter $\xi=E_0/(\omega c)$, where $c=1/\alpha\approx137$  is the speed of light. However, in SFPH experiments usually midinfrared light sources are used \cite{huismans2012} such that the recent improvements in the imaging techniques for photoelectrons \cite{eppink1997,doerner2000,hartung2019} offer the possibility to observe effects beyond the dipole approximation in SFPH. The transfer of photon momentum to the photoelectrons has already been observed in ``direct'', recollision-free ionization with circularly polarized drivers \cite{smeenk2011,hartung2019}. In addition, in linearly or slightly elliptically polarized laser pulses a counter-intuitive shift of the low-energy rescattered electrons against the light propagation direction \cite{ludwig2014,maurer2018} has been attributed to the interaction of the tunneled electron with the Coulomb potential. The structures have been analyzed by direct numerical solution of the TDSE \cite{chelkowski2015,ivanov2016} and have been interpreted in classical CTMC simulations \cite{tao2017} as well as semiclassical CCSFA calculations \cite{keil2017}. The theoretical analysis of the classical caustic \cite{danek2018} predicts that the shift of the beyond-dipole distribution depends on the momentum $p_x$ in polarization direction. However, as pointed out above, the caustic is an indication that the semiclassical analysis breaks down \cite{kleinert2009,xia2018}. 

In this paper, we investigate the photoelectron holography beyond the electric dipole approximation. To this end, we develop a semiclassical model based on the path integral approach \cite{shvetsov2016,lai2015} that includes beyond-dipole corrections properly in the associated phase as well as initial distribution of electron velocities. Using this semiclassical model, we can quantitatively explain the shift of the central holographic finger in 2D by analyzing the interference between two relevant trajectories. Physical insight is gained by a beyond-dipole simple man's model that allows a qualitative, analytical treatment. In 3D, the Coulomb focusing beyond the dipole approximation is analyzed in detail by  means of a glory rescattering model \cite{xia2018}. Our theory shows that the central maximum is dominated by the quantum interference of an infinite number of semiclassical paths and hence also in 3D the phase of the trajectories plays an important role. The high-order interference maxima in SFPH are also modified by nondipole effects which has been qualitatively observed by Chelkowski \textit{et al.} \cite{chelkowski2015}. Here, we investigate the shift of the interference structure for various orders of the holographic fringes and analyze also their forward/backward asymmetry in the emission strength. We compare all results to the full numerical solution of the nondipole TDSE in two and three dimensions.

\newpage

\section{Methods}
\subsection{A semiclassical model beyond the electric dipole approximation}
\label{sec:model}
Quantum mechanically, the photoelectron momentum distribution (PMD) is given by the modulus square of the overlap between the time-evolved state of the system $\psi(t_f)=U(t_f,t_0)\psi_0$ and the scattering state $\psi_{\textbf{p}}^{(-)}$ corresponding to an asymptotic momentum $\textbf{p}$:
\begin{equation}
M_{\textbf{p}}(t_f)=\langle \psi_{\textbf{p}}^{(-)}|U(t_f,t_0)|\psi_0\rangle.
\label{overlap}
\end{equation}
The non-relativistic time-evolution of the initial state is determined by the TDSE 
\begin{equation}
\I\partial_t\psi(t)=H\psi(t),
\label{eq:TDSE}
\end{equation}
with the Hamiltonian $H$ in Coulomb gauge given by 
\begin{equation}
H(\textbf{r},\hat{\textbf{p}},t)=\frac{1}{2}\left(\hat{\textbf{p}}+\textbf{A}(\eta)\right)^2+V(\textbf{r}).
\label{eq:hamilcoul}
\end{equation}
For a plane-wave laser pulse the introduced vector potential $\textbf{A}(\textbf{r},t)=\textbf{A}(\eta)$ depends only on the light-cone coordinate $\eta=t-z/c$. It is related to the the magnetic field by $\textbf{B}=\nabla\times \textbf{A}$ and to the electric field by $\textbf{E}=-\partial_t\textbf{A}$. Here, we denote the static ionic potential by $V(\textbf{r})$.

As shown in Refs. \cite{walser2003,spanner2003} the quality of the semiclassical approximation improves tremendously (compared to its application to the time-evolution operator in Eq. \eqref{overlap}), if the time-evolution operator is portioned such that: (i) initial state evolves in the field-free atomic potential given by $\psi_0(t')=U_0(t',t_0)\psi_0=\exp(\mathrm{i}I_p(t'-t_0))\psi_0$ with the ionization potential $I_p$; (ii) at a time $t'$ the electron is ``kicked'' by the interaction operator $H_I(t)=\hat{\textbf{p}}\cdot\textbf{A}(\eta)+\textbf{A}^2(\eta)/2$; (iii) it subsequently interacts with the laser field as well as the potential. Using the Dyson equation, the ionization amplitude reads
\begin{equation}
M_{\textbf{p}}(t_f)=-\I \int_{t_0}^{t_f} \mathrm{d}t'\, \langle \psi_{\textbf{p}}^{(-)}|U(t_f,t')H_I(t')|\psi_0(t')\rangle.
\end{equation}
For large final times $t_f$ the photoelectron is far away from the ionic core and we can approximate the scattering state $\psi_{\textbf{p}}^{(-)}$ by a plane wave with the same asymptotic momentum $\textbf{p}$. Introducing the closure relation $\boldsymbol{1}=\int \mathrm{d} \textbf{r}\, |\tilde{\textbf{r}}\rangle \langle \tilde{\textbf{r}}|$ the ionization amplitude can be written as 
\begin{equation}
\begin{aligned}
M_{\textbf{p}}(t_f)=&-\I \int_{t_0}^{t_f} \mathrm{d}t'\int \mathrm{d}\tilde{\textbf{r}}\\ & \langle \textbf{p}|U(t_f,t')|\tilde{\textbf{r}}\rangle\, \langle \tilde{\textbf{r}}|H_I(t')|\psi_0(t')\rangle.
\end{aligned}
\label{eq:resolution}
\end{equation}
We evaluate the  mixed position-momentum-space propagator $\langle \textbf{p}|U(t_f,t')|\tilde{\textbf{r}}\rangle$ within the semiclassical approximation of Feynman's path integral in Lagrangian formulation (see, e.g. Ref. \cite{kleinert2009} for a textbook treatment). The main idea of this approach is that the (classical) action in the Feynman propagator is large compared to the quantum action $\hbar=1$ such that the path integral over (in general nonclassical) paths is dominated by the region around the classical path and can be approximated by the saddle-point method. The classical reference paths $\textbf{r}_s(t)$ satisfy the classical equation of motion including electric and magnetic field inhomogeneities:
\begin{align}
	\dot{\textbf{k}}_s(t)=&-\nabla_\textbf{r} V(\textbf{r}_s(t))-\frac{\textbf{e}_z}{c}\, \textbf{E}(\eta_s(t))\cdot \dot{\textbf{r}}_s(t),\\
	\dot{\textbf{r}}_s(t)=&\, \textbf{k}_s(t)+\textbf{A}(\eta_s(t)),
	\label{eq:EOM}
\end{align}
that are written as Hamilton's equations with a canonical momentum $\textbf{k}_s(t)$. All classical paths that fulfill the initial and final boundary conditions,  $\tilde{\textbf{r}}=\textbf{r}_s(t')$ and $\textbf{p}=\textbf{k}_s(t_f)$, regardless of the auxiliary initial momentum $\tilde{\textbf{p}}=\textbf{k}_s(t')$ have to be taken into account.
Neglecting Maslov's correction, the semiclassical propagator reads as (up to an overall phase)
\begin{equation}
\begin{aligned}
\langle \textbf{p}|U(t_f,t')|\tilde{\textbf{r}}\rangle
\approx \frac{e^{\I\varphi}}{(2\pi)^{3/2}}\sum_{s}\sqrt{\mathrm{det}\left[\frac{\partial \textbf{k}_s(t')}{\partial \textbf{k}_s(t_f)}\right]}
e^{\I S(\textbf{k}_s,\textbf{r}_s)}.
\end{aligned}
\label{eq:divergence}
\end{equation}
The action associated to a path is given by \cite{miller1971,kleinert2009}
\begin{equation}
\begin{aligned}
S[\textbf{k},\textbf{r}]=&-\textbf{k}(t')\cdot \textbf{r}(t')\\
&-\int_{t'}^{t_f} \mathrm{d}t \left[\dot{\textbf{k}}(t)\cdot \textbf{r}(t)+H(\textbf{r}(t),\textbf{k}(t),t)\right].
\end{aligned}
\label{eq:action1}
\end{equation}
At each time $t'$ and for every initial position $\tilde{\textbf{r}}$ we would need to search in the three-dimensional space of initial momenta $\tilde{\textbf{p}}$ all possible classical trajectories. For simplification, in the spirit of the two-step model we would like to introduce a connection between the auxiliary momenta $\tilde{\textbf{p}}$ and the initial positions $\tilde{\textbf{r}}$ of the classical trajectories and hence to unravel the $\tilde{\textbf{r}}$-integration in Eq. \eqref{eq:resolution}. Even though there is no fundamental justification, this concept of a defined tunnel exit is well established in the semiclassical modeling of strong-field phenomena, compare for example  \cite{shvetsov2004,yan2010,landsman2015}.

In addition, we apply the saddle-point approximation to the time-integral and hence we only need to take into account a finite number of complex-valued initial times. To this end, we have to continue the classical paths into complex time $t'=t_0+\I\tau$. In the spirit of the complex-time quantum-orbit model \cite{smirnova2014}, we choose a path consisting of two sections: (i) one from the complex valued time $t'=t_\mathrm{i}$ down to the real axis, i.e. $t'\rightarrow \mathrm{Re}(t')$ which can be identified with the under-the-barrier tunneling process; (ii) the acceleration step identified with the real motion in real time from the time $\mathrm{Re}(t')=t_0$ where the electron is born in the continuum up to the final time $t_f$ where the electron reaches the detector. In order to simplify the calculation we neglect the potential $V$ during the first step of tunnel ionization. Hence, in first order of $1/c$ an auxiliary electron momentum $\tilde{\textbf{p}}'$ (not equal to the momentum in Eq. \eqref{eq:EOM}) exists that is conserved during the potential-free motion and would be equal to the final momentum, if the potential was also neglected in the second step. In this case, the EOM \eqref{eq:EOM} can be solved analytically in first order of $1/c$. By assuming real-valued trajectories in the second step, i.e. $\mathrm{Im}(\textbf{r}(t_0))=0$, and a vanishing real-part of the position at the complex-valued time $t_s'$, i.e. $\mathrm{Re}(\textbf{r}(t_s'))=0$, we fix the occurring integration constants. Under these assumptions the position of the tunnel exit $\textbf{r}_\mathrm{i}$ at time $t_0=\mathrm{Re}(t_s')$ is given by (to first order of $1/c$)
\begin{equation}
\begin{aligned}
\textbf{r}_{\mathrm{i}}=&\mathrm{Re}\left(\left(1+\frac{\tilde{p}'_z}{c}\right)\boldsymbol{\alpha}(t_0,t_\mathrm{i})+\frac{\tilde{p}'_z(t_i-t_0)}{c}\textbf{A}(t_\mathrm{i})\right)\\
+&\frac{\textbf{e}_z}{c}\mathrm{Re}\left(\tilde{\textbf{p}}'\cdot \boldsymbol{\alpha}(t_0,t_\mathrm{i})+\frac{1}{2}\alpha_2(t_0,t_\mathrm{i})\right)
\end{aligned}
\label{eq:ri}
\end{equation}
with the integrals $\boldsymbol{\alpha}(t,t')=\int_{t'}^t\mathrm{d}\tau \textbf{A}(\tau)$ and $\alpha_2(t,t')=\int_{t'}^t\mathrm{d}\tau \textbf{A}^2(\tau)$. Defining the quantity 
\begin{equation}
\textbf{v}(\tilde{\textbf{p}}',t)=\tilde{\textbf{p}}'+\textbf{A}(t)+\frac{\textbf{e}_z}{c}\left(\tilde{\textbf{p}}'\cdot\textbf{A}(t)+\frac{1}{2}\textbf{A}^2(t)\right)
\label{eq:saddle}
\end{equation}
which is the velocity of the electron if $z=0$, the corresponding initial velocity at the tunnel exit reads (to first order of $1/c$)
\begin{equation}
\begin{aligned}
\dot{\textbf{r}}_{\mathrm{i}}=\textbf{v}(\tilde{\textbf{p}}',t_0).
\end{aligned}
\label{eq:rip}
\end{equation}
The potential-free saddle-point equation for $t_s'$ is in first order of $1/c$ the same as in plain nondipole SFA \cite{brennecke2018a,walser2000}
\begin{equation}
\frac{\textbf{v}^2(\tilde{\textbf{p}}',t_s')}{2}+I_p=0.
\label{eq:saddleSFA}
\end{equation}
For the chosen time-integration contour, we can split the action into a complex-valued part under the barrier that is in first order of $1/c$ given by
\begin{equation}
S^0_{\downarrow}=I_pt_s'-\int_{t'_s}^{t_0}\mathrm{d}t\, \frac{\textbf{v}^2(\tilde{\textbf{p}}',t)}{2}
\end{equation}
and a real-valued part outside the barrier
\begin{equation}
\begin{aligned}
S_{\rightarrow}=&-\int_{t_0}^{\infty}\mathrm{d}t\,\biggl[\frac{1}{2}\dot{\textbf{r}}^2(t)+V(\textbf{r}(t))\\&-\textbf{r}(t)\cdot \nabla V(\textbf{r}(t))
-\textbf{z}(t)\cdot(\dot{\textbf{r}}(t)\times \textbf{B}(\eta(t)))\biggr].
\end{aligned}
\label{eq:semiphase}
\end{equation}
Since we have chosen the real part of the initial position to zero at the complex-valued time $t_s'$, we neglect the first phase term in Eq. \eqref{eq:action1}.
In contrast to the existing CCSFA beyond-dipole approximation \cite{keil2017} Eq. \eqref{eq:semiphase}  recovers the beyond-dipole SFA action \cite{he2017,walser2000,brennecke2018a} in the limit of vanishing binding potential $V$. For this purpose, the last term proportional to the magnetic part of the Lorentz force is important. Since by construction the trajectory and hence phase $S_{\rightarrow}$ are real-valued after the tunnel exit, the ionization probability associated with a certain trajectory is fully determined by the imaginary part $\mathrm{Im}(S^0_{\downarrow})$ corresponding to the tunnel motion. For linear polarization, the distribution of initial velocities is shifted by $\approx I_p/(3c)$ in propagation direction compared to the dipole approximation. According to tunneling theory this additional electron velocity is induced by the action of the magnetic part of the Lorentz force on the electron during its quantum-mechanical under-the-barrier motion \cite{klaiber2013}.

Summarizing this part, to calculate the amplitude of ionization into a final state with momentum $\textbf{p}$ one has to take the following steps: (i) Solve nondipole SFA saddle-point equation \eqref{eq:saddleSFA} for each auxiliary momentum $\tilde{\textbf{p}}'$ and hence establish a connection between the auxiliary momentum $\tilde{\textbf{p}}'$ and the initial conditions (compare Eqs. \eqref{eq:ri}, \eqref{eq:rip}) for the classical trajectories of the second step staring at time $t_0=\mathrm{Re}(t_s')$. (ii) Solve the  classical equations of motion (here done using Runge-Kutta-Fehlberg method) to find all possible auxiliary initial momenta $\tilde{\textbf{p}}'$ leading to a given final momentum $\textbf{p}$. We only include the most important kinds of trajectories in our model such that this inversion problem can be tackled by using a combination of the multi-dimensional Newton method and the conjugate gradient method. (iii)  Finally, the transition amplitude can be expressed in terms of these classical trajectories as
\begin{equation}
M_{\textbf{p}}(t_f)\propto  -\I\sum_{s} C(\tilde{\textbf{p}}',t_s) D(\tilde{\textbf{r}},t_s')\, \frac{1}{\sqrt{J}}\,
e^{\I (S^0_{\downarrow}+S_{\rightarrow})}.
\label{eq:finalamplitude}
\end{equation} 
To calculate the Jacobian $J$ in Eq. \eqref{eq:finalamplitude} we consider the variation of the final momenta $\textbf{p}$ with the auxiliary momentum $\tilde{\textbf{p}}'$ for fixed value of the exit point and ionization time $t_s'$ and denote it as:
\begin{equation}
J=\mathrm{det}\left[\frac{\partial \textbf{k}_s(t_f)}{\partial \textbf{k}_s(t_s')}\right]=\mathrm{det}\left[\frac{\partial\textbf{p}}{\partial \tilde{\textbf{p}}'}\right].
\label{eq:jacobian}
\end{equation} 
If the exit point as well as the time $t_s'$ are also varied with the momentum $\tilde{\textbf{p}}'$ according to Eqs. \eqref{eq:ri}, \eqref{eq:saddleSFA}, the calculated spectra are only weakly changed so that the conclusions of this paper stay the same.
In the actual implementation the pre-exponential factor $C$ and the matrix element $D(\tilde{\textbf{r}},t_s')=\langle\tilde{\textbf{r}}|H_I(t_s')|\psi_0(t_s')\rangle$ are neglected. 
We find that the leading-order corrections of the tunnel exit compared to dipole approximation have only negligible influence on the observables discussed below. However, the changed initial velocity distribution and hence the under-the-barrier action of the magnetic field are important for a successful, quantitative modeling of the signal strength.

\subsection{Numerical solution of the TDSE}
We benchmark our present semiclassical model against the exact numerical solution of the single-active electron time-dependent Schr\"odinger equation (TDSE) including leading-order nondipole corrections. We follow the scheme presented in Ref. \cite{brennecke2018} such that the theory covers the dynamics within electric quadrupole and magnetic dipole approximation. After application of a unitary transformation to the system in Coulomb gauge, compare Eqs. \eqref{eq:TDSE}, \eqref{eq:hamilcoul}, we obtain the numerical solution of the TDSE with a transformed Hamiltonian
\begin{equation}
\begin{aligned}
\widetilde{H}=&\frac{1}{2}\left(\textbf{p}+\textbf{A}(t)+\frac{\textbf{e}_z}{c}\left(\textbf{p}\cdot\textbf{A}(t)+\frac{1}{2}\textbf{A}^2(t)\right)\right)^2\\&+ V\left(\textbf{r}-\frac{z}{c}\textbf{A}(t)\right),
\end{aligned}
\label{eq:unihamiltonian}
\end{equation}
using the split-operator method on a Cartesian grid \cite{feit1983}. While propagating until the final time, outgoing parts of the wave function are projected onto beyond-dipole Volkov states \cite{brennecke2018a} and summed up coherently to obtain the momentum distribution \cite{lein2000}. Most of the results are calculated in reduced dimensionality (2D) with a soft-core model potential 
\begin{equation}
V_\textrm{2D}(r)=-\frac{e^{-0.575\, r}+1}{\sqrt{r^2+0.75}},
\end{equation}  
which mimics the helium atom with an ionization potential of $I_p\approx 0.9$~a.u. Selected results are also shown for a more realistic 3D calculation where the effective potential is chosen as by Tong \textit{et al}. \cite{tong2005}, but with the singularity removed using a pseudopotential \cite{troullier1991} for angular momentum $l=0$. In 2D, the size of the  numerical inner grid is $819$~a.u. in each direction with spacings of $\Delta x=0.2$~a.u. and a time step of $\Delta t=0.01$~a.u. is used. After the end of the pulse the simulation is run for five additional cycles. To obtain the PMD with high quality at low energies, we remove the localized bound states with a mask function ($r<30$~a.u.) from the final wave function and project the remaining wave function on eikonal states
\begin{equation}
\psi_{\textbf{p}}^{(e)}(\textbf{r})= \frac{1}{(2\pi)^{\frac{d}{2}}}\exp\left[\I\,\textbf{p}\normalsize{\cdot} \textbf{r}+\I \int_0^{\infty}\mathrm{d}\zeta\, V(\textbf{r}+\textbf{p}\zeta)\right].
\end{equation}
The momentum-space amplitude obtained from outgoing wave packets during the time-propagation and the parts projected onto eikonal states are added coherently to obtain the PMD with a resolution of $\Delta p_x=0.0038$~a.u. and $\Delta p_z=0.0019$~a.u. In 3D, the size of the  numerical inner grid is $358$~a.u. in each direction with spacings of $\Delta x=0.35$~a.u. and a time step of $\Delta t=0.02$~a.u. is used. In the plane spanned by the propagation and polarization direction, the PMD is obtained after the end of the pulse as in 2D with a resolution of $\Delta p_x=\Delta p_z=0.0044$~a.u.
\begin{figure*}
	\begin{center}
		\includegraphics[width=0.96\textwidth,angle=0]{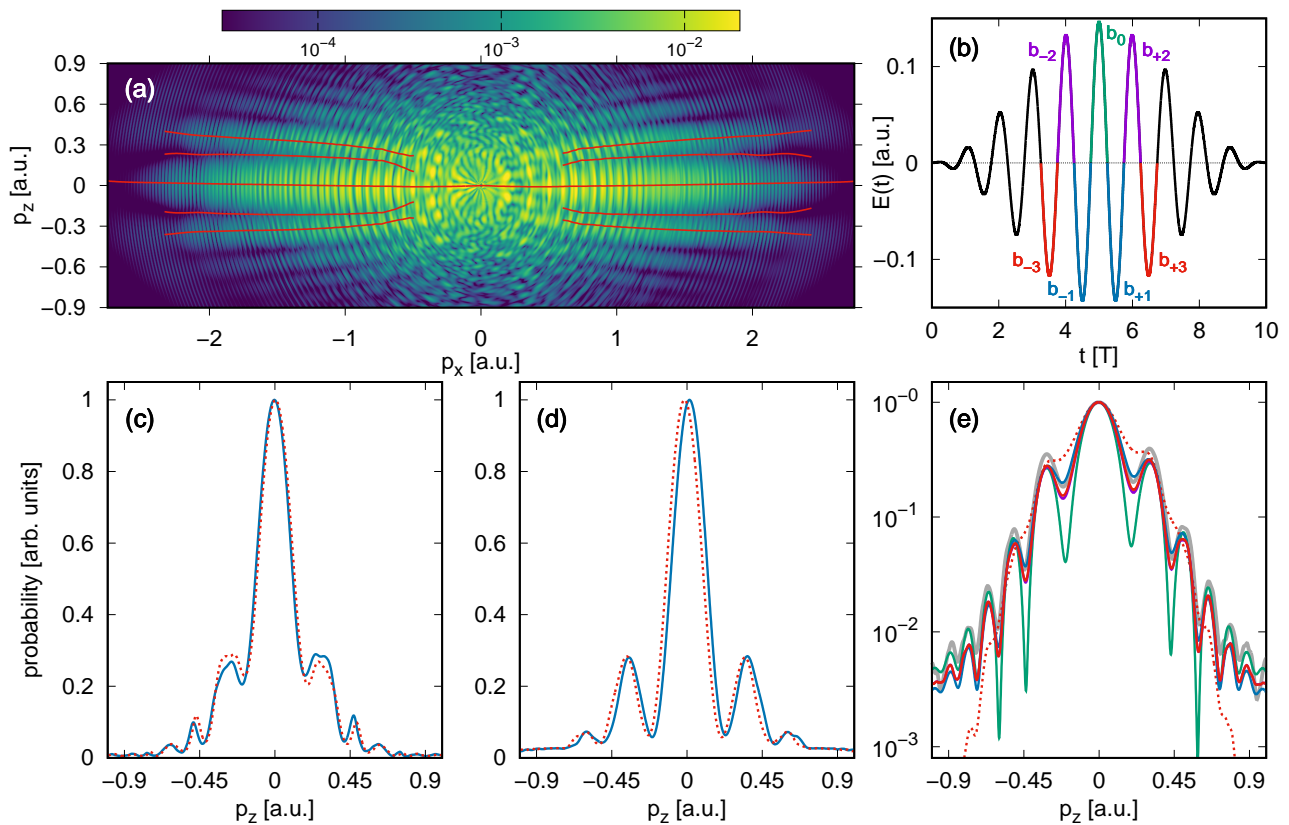}
		\vspace{-0.65cm}
	\end{center}
	\caption{(a) Intensity averaged photoelectron momentum distribution $w(\textbf{p})$ from ionization of helium  by a ten-cycle pulse with 800~nm wavelength and an average intensity of $7.5\times10^{14}$~W/cm$^{2}$, obtained by 2D TDSE simulation. The red lines are the numerically-determined positions of the holographic interference extrema. (b) Electric field with color-coded half cycles. (c),(d) 1D slices through the 2D distribution of panel (a) at fixed $p_x$: (c) $p_x=0.75$~a.u. and (d) $p_x=2.0$~a.u. (blue solid lines). To expose the nondipole shift and asymmetry, the mirror images, i.e. 1D slices through $w(p_x,-p_z)$, are  shown additionally (red dashed lines). (e) Comparison of the 1D slice at $p_x=1.0$~a.u. extracted from 2D TDSE (gray curve) to the semiclassical model (on a logarithmic scale): The colors indicate the included number of branches, compare panel (b). Only trajectories starting in the central half cycle $b_0$ are used for the green line; the blue line additionally take the branches $b_{\pm1}$ into account and so on. The dashed line corresponds to a calculation where the Jacobian $J$ instead of its square root is used, see main text for further information.}  
	\label{fig1}
\end{figure*}

\section{Results and discussion}
In our simulations we use a few-cycle linearly polarized laser pulse of $n_p$-cycle duration defined in terms of the vector potential 
\begin{equation}
\textbf{A}(\eta)=-\frac{E_0}{\omega}\sin^2\left(\frac{\omega\eta}{2n_p}\right)\sin(\omega \eta)\,\textbf{e}_x
\end{equation}
with a central frequency of $\omega=0.0569$~a.u. corresponding to $\approx800$~nm wavelength. For most calculations we use an intensity of $I=7.5\times10^{14}$~W/cm$^2$. However, to avoid the influence of intra-cycle interferences we additionally calculate ``intensity averaged'' spectra by summing up the PMDs corresponding to intensities $7\times10^{14}$, $7.5\times10^{14}$ and $8\times10^{14}$~W/cm$^2$. Even though the potentials used in the TDSE calculations are soft-core potentials with short-range contributions, we will use a bare Coulomb potential $V(r)=-1/r$ in the semiclassical calculations but set the ionization potential artificially to the helium value $I_p\approx0.9$~a.u.

\subsection{Comparison of TDSE and semiclassical results}
The photoelectron momentum distribution obtained by solution of the 2D TDSE for ionization of helium with a ten-cycle laser pulse is shown in Fig. \ref{fig1}(a). For the used intensity of $7.5\times10^{14}$~W/cm$^2$ the classical $2U_p$ cut-off for non-scattered electrons is given by $|p_x|\approx2.56$~a.u. The whole PMD is overlaid with ATI rings, i.e. peaks separated by the photon energy $\omega$, that result from inter-cycle interferences. For momenta $|\textbf{p}| \gtrapprox0.6$~a.u. pronounced holographic fingers that are nearly parallel to the polarization axis are visible. In contrast to the dipole limit, the exact positions of the minima and maxima indicated by the red solid lines are not symmetric with respect to the polarization axis \cite{chelkowski2015}. This symmetry breaking in propagation direction ($z$-direction) is more clearly visible in 1D slices through the distribution shown in panels (c) and (d). To suppress the strong oscillations from ATI rings, the slices are averaged over an interval of $\Delta p_x=0.1$ a.u. In addition to the shift of the peak positions, which are evident in panel (d), the emission strengths of the peaks in forward and backward directions are different, see panel (c).

The \textit{ab initio} PMD from the 2D TDSE solution can be interpreted within the  semiclassical model introduced in Section \ref{sec:model}. As in the saddle-point treatment of plain SFA, for a given auxiliary momentum $\tilde{\textbf{p}}'$ there is a single saddle-point time $t_s'$ in every half cycle of the laser pulse. Since the ionization weight $\exp(-\mathrm{Im}(S^0_{\downarrow}))$ depends strongly on the  electrical field strength and hence ionization time we will only consider trajectories that start in the vicinity of the maximum of the pulse envelope, compare the colormarked branches in Fig. \ref{fig1}(b).
For sufficiently long laser pulses the simple man's model predicts that trajectories starting in the ascending quarter cycles of the electric field strength do not return to the ionic core and hence are called ``direct'' electrons. In dipole approximation the potential-free mapping between the initial time $t_0$ and the final momentum $\textbf{p}$ is given by $\textbf{p}=-\textbf{A}(t_0)$. Hence, for the half plane with $p_x>0$ trajectories starting in the colored branches $b_{\pm1}$ and $b_{\pm3}$ of Fig. \ref{fig1}(b) are only weakly influenced by the Coulomb potential such that their Jacobian $J$ is close to unity. On the other hand, trajectories that start in the descending quarter cycle of the electric field strength are accelerated away and back to the parent ion such that they are strongly influenced by the potential. In our discussion we only include forward scattered electrons since other kinds of trajectories have much larger Jacobians $J$ and therefore only affect weakly the SFPH. Under this assumption, the main part of the holographic interference pattern is formed as follows. For a given final momentum $\textbf{p}$ there start two distinct trajectories in each of the colored branches $b_{0}$ and $b_{\pm2}$ of Fig. \ref{fig1}(b): one passing the parent ion with $z<0$ and the other with $z>0$. The interference of two such trajectories starting in the same quarter cycle (sometimes termed as type A holographic interference) leads to holographic fingers \cite{bian2011,shvetsov2018} which are nicely visible in a slice through the distribution at $p_x=1.0$~a.u. shown in Fig. \ref{fig1}(e).< Already a calculation that takes only trajectories from the central branch $b_0$ into account reproduces the right number of fringes and also their positions well. However, for a good agreement of the modulation depth the blue branches $b_{\pm1}$ of ``direct'' electrons have to be considered. Adding more branches only weakly affects the spectrum for $|p_x|\gtrapprox 0.5$~a.u. but lead to a complicated interference structure at low energies. 

Instead of solving the inversion problem, we can also calculate semiclassical PMDs by using the shooting method, i.e. a Monte-Carlo algorithm is used to sample a large number of trajectories corresponding to the possible auxiliary momenta $\tilde{\textbf{p}}'$ and finally the probability is obtained by binning (see Refs. \cite{shvetsov2016,keil2017} for a detailed discussion). As expected for the same initial conditions the resulting PMDs are (in the relevant momentum region) in agreement with Eq. \eqref{eq:finalamplitude}, if the Jacobian $J$ is used in Eq. \eqref{eq:finalamplitude} instead of its square root. This shows that all relevant branches have been included in the presented calculations using the inversion method. However, while our semiclassical model closely matches the TDSE holographic pattern and reproduces quantitatively the signal strength at larger transversal momenta $|p_z|$, the shooting-method results decrease asymptotically too fast, see Fig. \ref{fig1}(e).

\subsection{The central holographic fringe without Coulomb focusing}
\label{sec:shift}
In two-dimensional calculations and also for short-range potentials in three dimensions, Coulomb focusing is absent. As pointed out above, in these cases the central holographic finger is ``shaped'' out of the classical distribution by the constructive interference between forward scattered trajectories starting in the same quarter cycle. We have found nearly the same dependence of the nondipole shift on the momentum $p_x$ for the different settings (not shown). This indicates that in contrast to recent studies \cite{chelkowski2015,keil2017,danek2018} the nondipole modifications of the classical distribution cannot provide a complete and consistent interpretation of the momentum-dependent shift of the central maximum. In order to gain understanding of the influence of nondipole effects on the holographic interference, we first present the simplest possible model that allows an analytical treatment and offers a qualitative interpretation of the nondipole shifts. To this end we extend our classical model originally introduced in Refs. \cite{brennecke2018,brennecke2018a} for high-order above-threshold ionization to the case of photoelectron holography. 
We follow the ideas reported in Refs. \cite{bian2011,bian2012} and analyze the interference between a non-scattered reference wave and a scattered signal wave but include leading order nondipole effects properly.

After the ionization has launched an electron at the ionization time $t_0$ with an initial velocity $\textbf{v}_0\perp \textbf{e}_x$, the potential-free acceleration of the electron is described classically by Newton's equation. In leading order of $1/c$ we find a conserved canonical momentum of
\begin{equation}
\tilde{\textbf{p}}'=-\textbf{A}(t_0)+\frac{\textbf{A}^2(t_0)}{2c}\textbf{e}_z+\textbf{v}_0
\label{eq:momentum}
\end{equation}
in the transformed system defined by the Hamiltonian of Eq. \eqref{eq:unihamiltonian} without potential. In this model, the reference wave does not interact with the ionic potential and goes directly to the detector. Its canonical momentum $\tilde{\textbf{p}}'$ is conserved and the mapping between the final momentum $\textbf{p}$ and the initial time $t_0^{\mathrm{ref}}$ is given by Eq. \eqref{eq:momentum}. Using Eq. \eqref{eq:semiphase} the associated semiclassical phase is in leading order of $1/c$ given by 
\begin{equation}
S_\mathrm{ref}=-\int_{t_0^{\mathrm{ref}}}^\infty \mathrm{d}t\, \frac{\textbf{v}^2(\textbf{p},t)}{2}
\end{equation}
with the wave vector $\textbf{v}$ of Eq. \eqref{eq:saddle}.
On the other hand, the signal wave packet scatters off the parent ion. Since the magnetic part of the Lorentz force causes a drift motion of the electron in propagation direction, for an exact return to the initial position $\textbf{r}=0$ at the recollision time $t_r$ the electron has to start with an initial velocity $\textbf{v}_0=-|v_0|\textbf{e}_z$ against the propagation direction of the light \cite{walser2000}:
\begin{equation}
v_{0,z}=-\frac{1}{2c(t_r-t_0^{\mathrm{sig}})}\int_{t_0^{\mathrm{sig}}}^{t_r}\mathrm{d}\tau\, (\textbf{A}(\tau)-\textbf{A}(t_0^{\mathrm{sig}}))^2.
\label{eq:vz0}
\end{equation}
The recollision time $t_r$ can be found from the return condition in polarization direction, $x(t_0^{\mathrm{sig}})=x(t_r)$, that is the same as in the dipole limit. During the scattering the electron feels the potential and its canonical momentum changes from $\tilde{\textbf{p}}'$ of Eq. \eqref{eq:momentum} to the final momentum $\textbf{p}$. 
\begin{figure}[t]
	\begin{center}
		\includegraphics[width=1.0\columnwidth,angle=0]{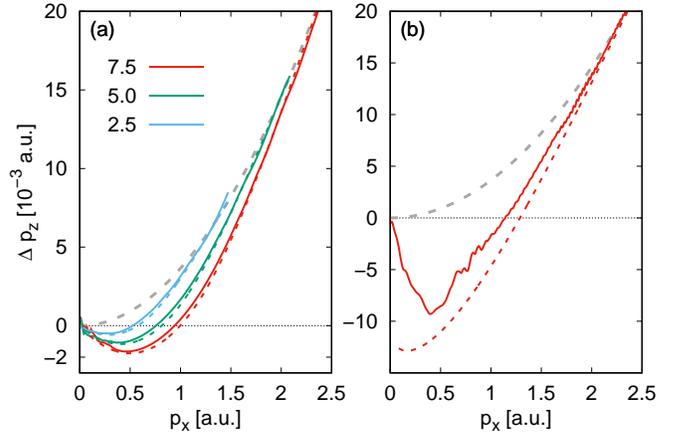}
		\vspace{-1.00cm}
	\end{center}
	\caption{Shift of the central holographic finger extracted from 2D TDSE simulations (solid lines): (a) Results for various intensities, given in the legend in units of $10^{14}$ W/cm$^2$, for a two-cycle pulse with 800~nm wavelength; (b) intensity-averaged result for a ten-cycle pulse, see Fig. \ref{fig1}. The TDSE results are compared to the simple man's model result given by Eq. \eqref{eq:shiftmodel} (colored dashed lines). The gray dashed line represents the classical estimate assuming ``direct'' ionization: $\Delta p_z=p_x^2/(2c)$, see Eq. \eqref{eq:momentum} with $\textbf{v}_0=0$.}   
	\label{fig2}
\end{figure}
For fixed times $t_0^{\mathrm{sig}},t_r$ the energy conservation during rescattering implies that the possible outgoing velocities $\textbf{v}(\textbf{p},t_r)$ after the scattering lie on a circle of radius $K(t_0^{\mathrm{sig}},t_r)\approx |\textbf{A}(t_r)-\textbf{A}(t_0^{\mathrm{sig}})|$ that is in leading order of $1/c$ the same as in the dipole approximation \cite{brennecke2018}. After the recollision the second acceleration stage maps the outgoing velocity to the final velocity that is equal to the canonical momentum $\textbf{p}$. Therefore, the phase of the signal electron is given by
\begin{equation}
S_\mathrm{sig}=-\int_{t_0^{\mathrm{sig}}}^{t_r} \mathrm{d}t\, \frac{\textbf{v}^2(\tilde{\textbf{p}}',t)}{2}-\int_{t_r}^{\infty} \mathrm{d}t\, \frac{\textbf{v}^2(\textbf{p},t)}{2}.
\end{equation}
The interference pattern is determined by the acquired phase difference between the reference and the signal wave packets that reads as 
\begin{equation}
\Delta S=\int_{t_0^{\mathrm{ref}}}^{t_r} \mathrm{d}t\, \frac{\textbf{v}^2(\textbf{p},t)}{2}-\int_{t_0^{\mathrm{sig}}}^{t_r} \mathrm{d}t\, \frac{\textbf{v}^2(\tilde{\textbf{p}}',t)}{2}.
\label{eq:phasediff}
\end{equation}
The holographic fingers are caused by interference of trajectories starting in the same quarter cycle of the field such that (close to the polarization axis) the starting times of the signal and reference wave, $t_0^{\mathrm{sig}}$ resp. $t_0^{\mathrm{ref}}$, are quite similar.
For this type of interference the $z$-component of the first term of Eq. \eqref{eq:phasediff} can be identified as the dominant term such that in first order of $1/c$ we can approximate
\begin{equation}
\Delta S \approx \int_{t_0^{\mathrm{ref}}}^{t_r} \mathrm{d}t\, \left[\frac{p_z^2}{2}+\frac{p_z}{c}\left(\textbf{p}\cdot\textbf{A}(t)+\frac{1}{2}\textbf{A}^2(t)\right)\right].
\label{eq:difphase}
\end{equation} 
In dipole approximation, the positions of the extrema are determined by the first term of Eq. \eqref{eq:difphase}. Since the nondipole shifts are of the order of $1/c$ and we concentrate on the holographic fingers close to the polarization axis, we assume the same initial and return times as in dipole approximation. The positions of the holographic maxima are prescribed by points of constructive interference in Eq. \eqref{eq:phasediff}: $\Delta S=2\pi n$ with $n\in \mathbb{Z}$. The central finger corresponds to forward scattering, i.e. $\tilde{\textbf{p}}'=\textbf{p}$, so that (in this simple man's picture) the signal and the reference trajectory get equivalent. As a result, the position of the central lobe is determined by the interplay between needed initial velocity $\textbf{v}_0$ and the momentum $\textbf{A}^2(t_0^\mathrm{sig})/(2c)$ transfered from the electromagnetic field to the electron during its acceleration. The zero of the phase difference $\Delta S$ of Eq. \eqref{eq:phasediff} is at the same position as its minimum what can be used to calculate the shift. On the other hand, to estimate the positions of the higher-order extrema we expand the approximated phase difference of Eq. \eqref{eq:difphase} in first order of $1/c$ around the dipole position $p_z^{\mathrm{dip}}$ and solve analytically the resulting equation for the shift $\Delta p_z$. Under these assumptions, we find for the central lobe as well as all side lobes in first order of $1/c$ the same form of the shift
\begin{equation}
\Delta p_z \approx -\frac{1}{c}\frac{1}{t_r-t_0^{\mathrm{ref}}} \int_{t_0^{\mathrm{ref}}}^{t_r} \mathrm{d}t\, \left(\textbf{p}\cdot\textbf{A}(t)+\frac{1}{2}\textbf{A}^2(t)\right).
\label{eq:shiftmodel}
\end{equation}
Close to the classical cut-off $p_x\approx -A_x(t_0^{\mathrm{ref}})\approx -E_0/\omega$ the excursion time $t_r-t_0^{\mathrm{sig}}\approx t_r-t_0^{\mathrm{ref}}$ of the electron and the corresponding initial velocity $\textbf{v}_0$ of Eq. \eqref{eq:vz0} vanish such that the shift of Eq. \eqref{eq:shiftmodel} simplifies $\Delta p_z\approx A^2(t_0)/(2c)=p_x^2/(2c)$. 
\begin{figure}[t]
	\begin{center}
		\includegraphics[width=1.0\columnwidth,angle=0]{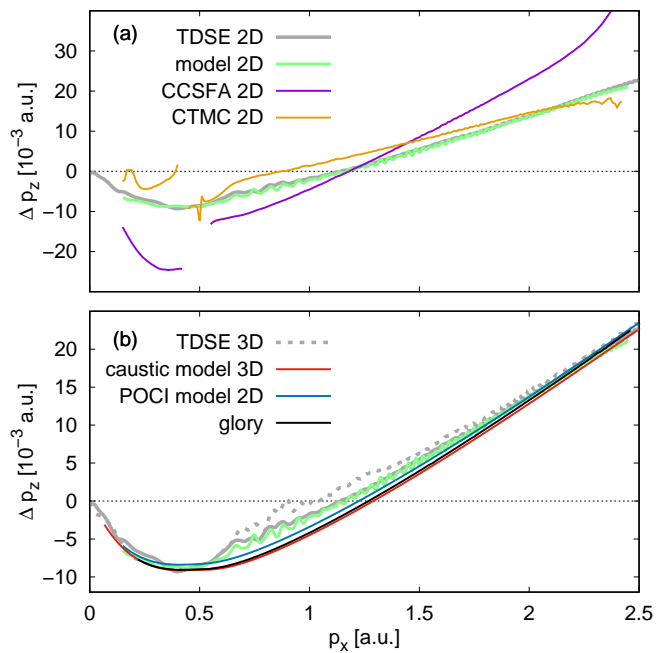}
		\vspace{-1.00cm}
	\end{center}
	\caption{Comparison of the shift of the central maximum obtained by different stages of theory under the conditions as in Fig. \ref{fig1}. The position of the maximum extracted from TDSE simulations in 2D and 3D are given by gray solid and dashed lines, respectively. The position of the maximum of our semiclassical model in 2D including the branches $b_0$, $b_{\pm1}$ is shown as green line and the corresponding (numerically-determined) point of constructive interference (POCI) for the two trajectories starting in branch $b_0$ as blue line. The yellow line corresponds to the maximum of the classical CTMC distribution in 2D whereas the red line belongs to the position of the (numerically-determined) 3D caustic. The maximum of the glory model $\delta p_z$ is given by the black line.}  
	\label{fig3}
\end{figure}
Hence, the shift is fully determined by the acceleration stage and it is the same as in the simple man's model for direct ionization, compare Eq. \eqref{eq:momentum}. On the other hand, for low momenta $p_x\approx0$ the formula predicts negative values that can be estimated by $\Delta p_z\propto -U_p/c$ for long laser pulses.
Since low-energy electrons dominate the position of the maximum of the lateral distributions integrated over $p_x$, we can explain the linear scaling of the backward shift with intensity as observed in Ref. \cite{chelkowski2015} for the solution of the 2D TDSE at $3400$~nm wavelength. 

To obtain a reliable numerical value that quantifies the shift of the central fringe for TDSE results, we first average the PMDs over an interval of $\Delta p_x=0.1$~a.u. to suppress the influence of ATI peaks. Afterwards we perform Gaussian fits to the central region with $|p_z|<0.02$~a.u. at each longitudinal momentum $p_x$ from which the position of the maximum is calculated. For very short two-cycle pulses where classically the electrons can undergo only a single forward scattering, the simple formula of Eq. \eqref{eq:shiftmodel} predicts perfectly the momentum-dependent shift of the central finger for different intensities, see Fig. \ref{fig2}(a). For this special pulse shape the rescattering electrons with small final momenta $p_x$ recollide after the end of the laser pulse. Hence, the denominator of Eq. \eqref{eq:shiftmodel} becomes large, so that the shift is small at low energies. In long laser pulses, the agreement between our simple model and the TDSE is still good for high-energies, see Fig. \ref{fig2}(b). 
At low energies, however, higher-order rescatterings become important \cite{hickstein2012,liu2011,kästner2012} which also modify the nondipole dynamics \cite{danek2018} but are not incorporated in our simple model. Here, the semiclassical model developed in Section \ref{sec:model} can provide a quantitative interpretation of the shift, see Fig. \ref{fig3}(a). The TDSE as well as the semiclassical model (in 2D) predicts a minimal shift at $p_x\approx0.45$~a.u. For larger longitudinal momenta $p_x$ only a single recollision exists and the agreement between the semiclassical model and the TDSE calculation is perfect. On the other hand, for small $p_x$ the shift goes to zero which has been classically attributed to multiple recollisions~\cite{danek2018}. 

While our semiclassical model agrees quantitatively with the TDSE result, the earlier proposed CCSFA beyond the dipole approximation \cite{keil2017} overestimates the magnitude of the shift, which is caused by the different phase in CCSFA that can be obtained from Eq. \eqref{eq:semiphase} by omitting the terms $-\textbf{r}\cdot \nabla V(\textbf{r})$ and $-\textbf{z}\cdot(\dot{\textbf{r}}\times \textbf{B})$ in the integrand. In additional calculations (not shown) we found that both terms are important for the good agreement of our model. If we neglect the phase and use the classical trajectory Monte Carlo (CTMC) distribution in 2D, we obtain a much to broad maximum whose position cannot quantitatively reproduce the TDSE results. Hence, in contrast to the statements in Refs. \cite{chelkowski2015,keil2017}, in 2D the position of the central lobe cannot be explained by the nondipole modifications of the classical trajectories alone. Instead the point of constructive interference (POCI) of the two trajectories of branch $b_0$ determines the position of the central lobe, see Fig. \ref{fig3}(b). Compared to the simple man's model their phase difference in the semiclassical model varies linearly around the position of the central maximum and has a first-order zero. The small difference between the POCI and shift of the full semiclassical calculation is mostly caused by the presence of ``direct'' electrons starting in branches $b_{\pm1}$. The saddle-point treatment of nondipole SFA shows that in recollision-free ionization with linearly polarized fields the lateral distributions is approximately shifted in propagation direction by $\Delta p_z=p_x^2/(2c)+I_p/(3c)$. Hence, the position of the maximum of the complete distribution is slightly shifted to larger $p_z$ compared to the POCI.

\subsection{The central holographic fringe with Coulomb focusing}
\begin{figure}[t]
	\begin{center}
		\includegraphics[width=1.0\columnwidth,angle=0]{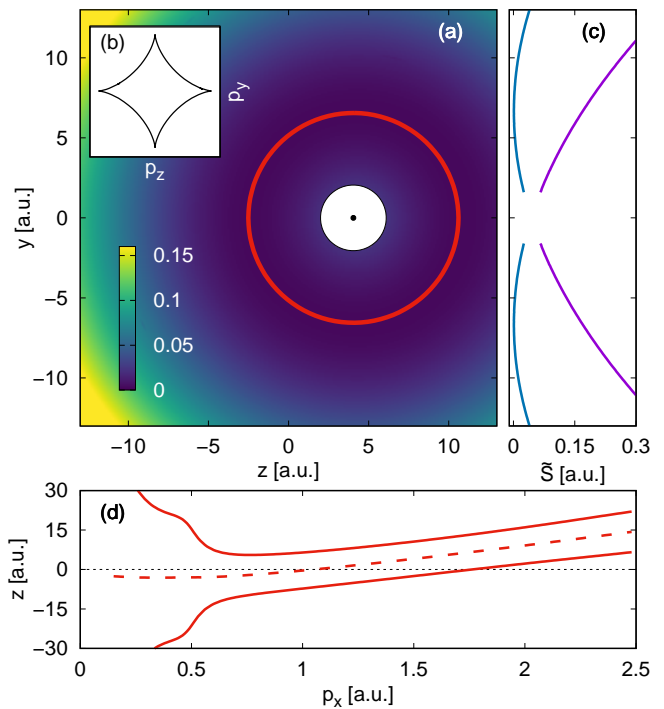}
		\vspace{-1.00cm}
	\end{center}
	\caption{(a) The transformed phase $\tilde{S}$ of Eq. \eqref{eq:newphase} in $y$-$z$-space for fixed $p_x=1.5$~a.u. at the end of the laser pulse (shifted to $\tilde{S}=0$ at the minimum). The red line indicates the positions of trajectories with vanishing Jacobian $J$ that correspond to the caustic structure in final-momentum space shown schematically in the inset (b). (c) Slice trough the phase-map $\tilde{S}$ of Eq. \eqref{eq:newphase} for fixed $z=z_{p_x}^0$. The blue line corresponds to the branch shown in panel (a) whereas the violet line represents the other important branch for the formation of holographic fingers. (d) Slice at $y=0$ through the tube with vanishing Jacobian $J$ in mixed-coordinate representation (solid lines) where the dashed line indicates the center of the tube.}  
	\label{fig4}
\end{figure}

In three dimensions, the two-path interference introduced above still determines the holographic finger pattern for large lateral momenta $p_{\perp}$. Near the polarization axis, however, the picture is modified by Coulomb focusing.
In dipole approximation, the cylindrical symmetry implies that on the polarization axis an  infinite number of distinct classical trajectories lead to the same final momentum $\textbf{p}$. In this picture, a circle of initial transverse velocities at the tunnel exit is mapped to one point in the final momentum distribution such that in the dipole limit an axial caustic singularity emerges, characterized by a vanishing Jacobian $J$ of Eq. \eqref{eq:jacobian}. Hence, the PMD scales with $1/p_\perp$ close to the polarization axis. Even beyond the dipole approximation, Maurer \textit{et al.} and Dan\ifmmode \check{e}\else \v{e}\fi{}k \textit{et al.} \cite{maurer2018,danek2018} observed a divergent caustic structure close to the polarization axis. In the space of auxiliary momenta $\tilde{\textbf{p}}'$ the zeros of the Jacobian $J$ form a tube that is shifted against the light propagation direction (compared to the rotationally symmetric dipole limit) to compensate the magnetically induced drift motion, compare Eq. \eqref{eq:vz0}. In contrast to the dipole approximation, for a given final longitudinal momentum $p_x$ the classical mapping leads to a complicated one-dimensional manifold of final transversal momenta $\{p_y,p_z\}$ with corresponding vanishing Jacobian $J$, see Fig. \ref{fig4}(b). 
However, since the width of this structure scales in leading order with $1/c^2$, it is negligible in the weakly relativistic region, even though it changes the type of the caustic.

The position of the classical caustic has been estimated in an analytical model in Refs. \cite{maurer2018,danek2018}. Interestingly, assuming only a single recollision of the electron, their result is the same as our position of the central maximum estimated within the simple man's model of Eq. \eqref{eq:shiftmodel}. Indeed in our semiclassical calculations the phases associated with the different trajectories corresponding to the caustic show only a minor variation for fixed momentum $p_x$.  Hence, we find numerically in the semiclassical model that the point of constructive interference (POCI) in 2D and the center of the classical caustic in 3D do not coincide exactly. Nevertheless, for all used laser parameters and target atoms both positions are quite close, see Fig. \ref{fig3}(b). We want to emphasize that their difference scales in leading order with $1/c$ so that even in the weakly relativistic regime they are not exactly the same. 

In the derivation of our semiclassical model, we used the semiclassical approximation of the propagator $\langle \textbf{p}|U(t_f,t')|\tilde{\textbf{r}}\rangle$ that is based on a stationary-phase approximation to the corresponding path integral. The critical ``points'' of the action are the solutions of the classical equations of motion. The second ``variation'' of the action is related to the Jacobian $J$ of the system such that a vanishing Jacobian $J$ and the appearance of caustics indicates the breakdown of the semiclassical approximation \cite{dangelmayr1979,dewittmorette1997}. Recently, a regularization procedure has been applied to Coulomb focusing in strong-field ionization in dipole approximation \cite{xia2018}. There, the similarity of laser-induced rescattering to glory scattering \cite{berry1969} is used and an analogous procedure as worked out in Ref. \cite{dewittemorette1984} is applied to obtain finite spectral weights close to the classical singularity. The main idea is that even though the semiclassical approximation of the propagator $\langle \textbf{p}|U(t_f,t')|\tilde{\textbf{r}}\rangle$ is singular for certain trajectories, the semiclassical approximation of the propagator $\langle \textbf{r}|U(t_f,t')|\tilde{\textbf{r}}\rangle$ can be finite for the same trajectories \cite{littlejohn1992}. Hence, instead of directly calculating the amplitude $M_\textbf{p}(t_f)$ we first determine the position representation of the outgoing photoelectron wave packet $M_\textbf{r}(t_f)$ by using the semiclassical approximation and afterwards apply a Fourier transformation to obtain the PMD:
\begin{equation}
M_\textbf{p}(t_f)=\frac{1}{(2\pi)^{3/2}}\int\mathrm{d}\textbf{r}\, M_\textbf{r}(t_f) e^{-\I\textbf{p}\cdot\textbf{r}}.
\label{eq:fourier}
\end{equation}
In the semiclassical treatment of $ M_\textbf{r}(t_f)$ we find, similar as with  $ M_\textbf{p}(t_f)$, for each position $\textbf{r}$ two relevant trajectories that start in the central half cycle $b_0$. For final $y>0$ and small $z$, one trajectory passes the parent ion in the range  $y>0$, corresponding to the blue curve in Fig. \ref{fig4}(c), and the other in the range $y<0$, corresponding to the violet curve. If we apply the steepest-descent method to integrals over $\textbf{r}$ in Eq. \eqref{eq:fourier}, we retrieve our divergent semiclassical model of Eq. \eqref{eq:finalamplitude}. It is still valid to apply the saddle-points method to the $x$-integration. However, the remaining integrals have to be treated more carefully. To this end, we will change of coordinates such that also in the nondipole regime the integrand of Eq. \eqref{eq:fourier} has an approximate rotational symmetry and resulting integrals can be simplified analytically.  

We find numerically that for a given finite final time $t_f$ and fixed $p_x$, the trajectories with vanishing Jacobian $J$ of Eq. \eqref{eq:jacobian} (whose final momenta $\textbf{p}$ belong to the complicated structure shown in Fig. \ref{fig4}(b)) form approximately in the $y$-$z$-plane a circle with radius $r_{p_x}^0$ and with its center shifted in propagation direction to $z_{p_x}^0$, see the red line in Fig. \ref{fig4}(a). A slice at $y=0$ through this tube is shown in Fig. \ref{fig4}(d) as a function of $p_x$ to illustrate the momentum dependence of the center shift $z_{p_x}^0$.
Hence, for each $p_x$ we shift the origin of our coordinate system to the center of the circle $z_{p_x}^0$ and introduce polar coordinates $r,\phi$:
\begin{equation}
y=r\sin(\phi)=y',\; z=z_{p_x}^0+r\cos(\phi)=z_{p_x}^0+z'.
\end{equation}
In addition, we write the final momenta as 
\begin{equation}
p_y=p_y',\;p_z=p_z'+\delta p_z.
\end{equation} 
Originating from the phase of Eq. \eqref{eq:action1} we define a new phase 
\begin{equation}
\tilde{S}=\mathrm{Re}(S^0_{\downarrow})+S_{\rightarrow}+\textbf{k}_s(t_f)\cdot(\textbf{y}_s(t_f)+\textbf{z}_s(t_f))-\delta p_z z',
\label{eq:newphase}
\end{equation}
which determines classically the associated momentum field \cite{littlejohn1992}, e.g. $p_y'=\partial \tilde{S}/\partial y'$. Since we are interested in small lateral momenta $p'=\sqrt{p_y'^2+p_z'^2}$, this motivates us to take only the branch with a minimum of $\tilde{S}$ into account, see Fig. \ref{fig4}(c). Using Fourier analysis we can chose $\delta p_z$ such that the phase $\tilde{S}$ is nearly constant for fixed $r$ as shown in panel (a) of Fig. \ref{fig4}. The remaining phase variation at constant $r=r_{p_x}^0$ is smaller that $10^{-3}$. Under these assumptions, the electron amplitude of Eq. \eqref{eq:fourier} can be simplified to
\begin{equation}
M_{\textbf{p}}(t_f)\propto \int \mathrm{d}rr \int \mathrm{d}\phi\, P(p_x,r,\phi)\, e^{\I \tilde{S}}e^{-\I(p_y'y'+p_z'z')},
\label{eq:specglory}
\end{equation}      
where the pre-factor $P$ contains all remaining real-valued components such as the ionization weights. Since for radii close to $r_{p_x}^0$ the pre-factor $P$ of Eq. \eqref{eq:specglory} is also only weakly angle dependent, the $\phi$-integration of the last plane-wave factor of Eq. \eqref{eq:specglory} is proportional to the zeroth-order Bessel-function $J_0(rp')$. 
For small $p'$ the $r$-integration can be carried out by saddle-point approximation which leads to the glory condition for the impact parameter $r_g$
\begin{equation}
\frac{\partial \tilde{S}(r)}{\partial r}\bigg|_{r=r_g}=0.
\end{equation} 
The resulting glory impact parameter $r_g$ is nearly the same as the radius $r_{p_x}^0$, schematically illustrated in Fig. \ref{fig4}(a). We want to emphasize that the above derivation is only applicable close to the position of the maximum: (i) for larger transversal momenta $p'$ the other branch of trajectories is non-negligible; (ii) around $z\approx z_{p_x}^0$ the pre-factor $P$ is singular which has to be treated properly for larger $p'$. 

The central maximum in the PMD is dominated by the quantum interference of an infinite number of semiclassical paths that give rise to a pattern of:
\begin{equation}
w(\textbf{p})\propto r_g J_0^2\left(r_g\sqrt{p_y^2+(p_z-\delta p_z)^2}\right)e^{-2\mathrm{Im}(S^0_{\downarrow})}.
\end{equation}
As a result, the position of the maximum is simply given by $\delta p_z$. The predicted $p_x$-dependence of this shift is nearly the same as for the position of the caustic as well as the semiclassical POCI in 2D, compare Fig. \ref{fig3}(b). For momenta lower than $p_x\approx0.45$~a.u. these theories agree perfectly with the exact result extracted from a slice with fixed $p_y=0$ through the 3D PMD obtained by numerical solution of the TDSE for helium. Similar as in the 2D results, the deviations around $p_x\approx0.8$~a.u. are caused by ``direct'' electrons from branches $b_{\pm1}$ (those that do not turn around on their way out). Since in 3D the Coulomb-focused maximum is flatter than in 2D, the ``direct'' electrons influence the position of the maximum more strongly. In contrast to the position of the classical caustic, this effect can be seen in the glory model with its finite spectral weight (not shown). 


\subsection{Higher-order holographic fringes}
Going to larger final lateral momenta $p_z$ the two interfering trajectories of the semiclassical model in 2D  become inequivalent in the sense that one is only weakly perturbed by the potential whereas the other is strongly deflected. Hence, the simple man's model with one signal and one reference beam becomes applicable and the interpretation of the results is easier. However, the identification of the positions of the extrema in real spectra from TDSE or semiclassical calculations requires care. To this end, we use the smoothed PMDs and perform  at each $p_x$ a Gaussian fit in the region around a maximum to determine its position. The shift of each order maxima is quantified by 
\begin{equation}
\Delta p_z(p_x)=\frac{p_z^++p_z^-}{2},
\end{equation} 
where $p_z^{+}$ and $p_z^{-}$ are the positions of one maximum in forward direction and one in backward direction, respectively. The extracted shift is then averaged over the ATI peaks to obtain the shown results.
\begin{figure}[b]
	\begin{center}
		\includegraphics[width=0.9\columnwidth,angle=0]{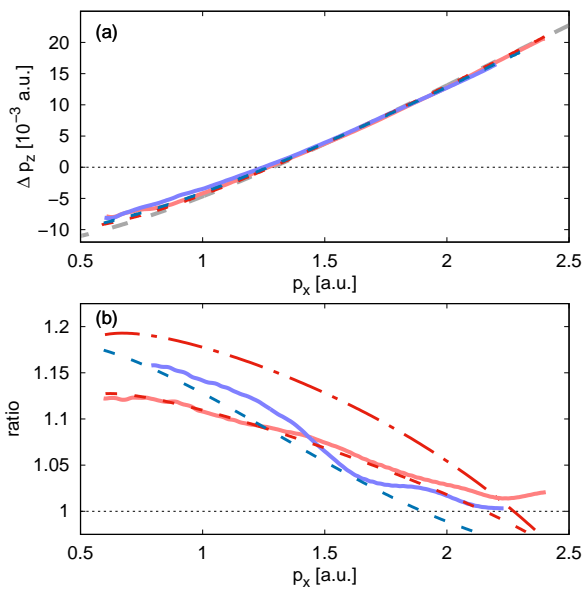}
		\vspace{-0.60cm}
	\end{center}
	\caption{Results for the first-order (red lines) and second-order (blue lines) maxima. (a) Shift of the holographic fringes.  (b) Forward/backward asymmetry quantified as the ratio of the signal strengths of the interference peaks. Solid lines, numerical TDSE results; dashed thin lines, results from the semiclassical model including only trajectories starting in branch $b_0$; gray dashed line, simple man's model given by Eq. \eqref{eq:shiftmodel} and dashed-dotted line semiclassical calculation using the Jacobian $J$ instead of its square root.}  
	\label{fig6}
\end{figure}
In Fig. \ref{fig6}(a) the momentum shifts $\Delta p_z$ of the first- and second-order maxima  extracted from 2D TDSE calculations are compared to the numerically obtained positions of the semiclassical model and the simple man's model of Eq. \eqref{eq:shiftmodel}.
The shift is nearly independent on the order of the holographic fringes and hence in the shown momentum region they are similar to the shift of the central maximum. This implies that for small $p_x$, backward shifts are observed. All levels of theory are in good agreement which indicates that the shift of the extrema can be entirely attributed to the phase difference between signal and reference trajectory starting in branch~$b_0$. 

On the other hand, the forward/backward asymmetry is quantified at each $p_x$ by taking the ratio of the signals of one interference maximum in forward direction and one in backward direction. In the relevant momentum region the 2D TDSE calculation shows always ratios larger unity indicating stronger emission in forward than in backward direction, see Fig. \ref{fig6}(b). The TDSE as well as semiclassical results show for high energies a stronger asymmetry for lower interference orders. 
\begin{figure}[t]
	\begin{center}
		\includegraphics[width=1.0\columnwidth,angle=0]{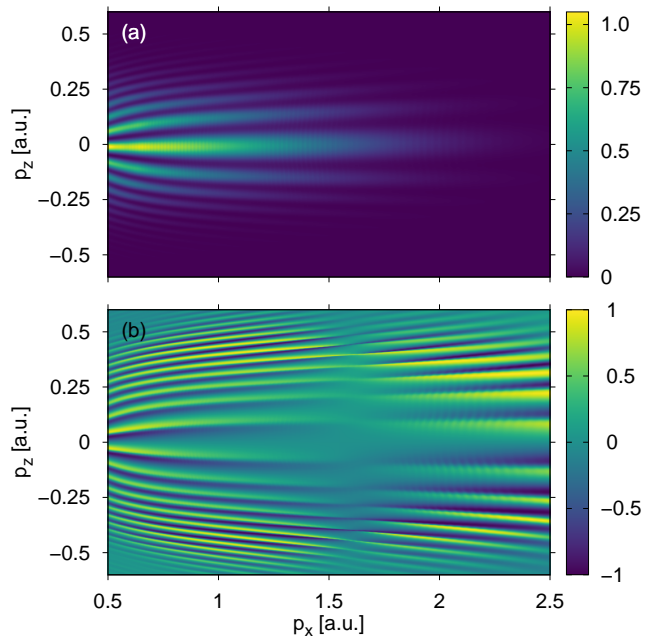}
		\vspace{-1.00cm}
	\end{center}
	\caption{(a) Smoothed photoelectron momentum distribution from ionization of a 2D soft-core hydrogen model $V=-1/\sqrt{r^2+0.64}$ by a three-cycle pulse with 3400~nm and an intensity of $5\times10^{13}$ W/cm$^2$. (b) Corresponding normalized difference $\mathrm{ND}$.}  
	\label{fig7}
\end{figure}
However, when going to smaller $p_x$ this changes such that around $p_x\approx0.8$~a.u. the asymmetry of the second-order maxima reaches values as high as $\approx1.17$. In the semiclassical calculation already the two dominating trajectories starting in branch $b_0$ can quantitatively reproduce the asymmetry of the first interference orders. In addition, we find that including trajectories from branches $b_{\pm1}$ improves slightly the agreement for the second orders for intermediate momenta $p_x$ (not shown). 
By using the Jacobian in Eq. \eqref{eq:finalamplitude} instead of its square root we can change the ``cross section'' used in the model and we find that the correct implementation is important for a quantitative modeling, see the dashed-dotted line in Fig. \ref{fig6}(b). In comparison with high-order above threshold ionization (HATI) \cite{brennecke2018a} we think that in the simple man's picture the asymmetry can be understood with the help of the $z$-component of the in-coming velocity shortly before rescattering and hence a rotated cross section compared to the symmetric dipole limit. However, in SFPH the recollision is much softer than in HATI such that the simple assumption of one hard rescattering event is not justified. For example, in contrast to HATI the distribution of initial velocities at the tunnel exit influences the asymmetry. If we take the distribution in dipole approximation and hence neglect the action of the laser magnetic field on the electron during its under-the-barrier motion \cite{klaiber2013}, we find that the asymmetry ratios decrease by $\approx0.03$ for the first-order holographic maxima. Close to the classical $2U_p$ cut-off, our semiclassical description using Coulomb-free trajectories during the under-the-barrier motion fails so that the spectral weight is not correctly modeled \cite{keil2016} and the asymmetry decreases too fast. 

The nondipole shift of the fringes scales quadratically with the electric-field strength $E_0$ and the  wavelength $\lambda$ (assuming that the momentum $p_x$ is scaled naturally as $p_x\propto E_0\lambda$). On the one hand, the spacing of the holographic interference fringes decreases with increasing wavelength. Even though it is well known that Coulomb effects influence strongly the exact positions of the fringes, we can estimate the wavelength dependence of the fringe spacing by simple man's model to $\propto \lambda^{-1/2}$. On the other hand, for sufficiently long wavelength the excursion time $t_r-t_0^\mathrm{sig}$ of the signal electron and hence the fringe spacing is nearly independent of the field strength $E_0$ \cite{huismans2012}. 
As a result for strong mid-infrared laser pulses as often used in today's experiments, the nondipole shift and the fringe spacing are in the same order of magnitude, see the momentum distribution in Fig. \ref{fig7}(a). In this parameter region, the huge influence of nondipole effects can be beautifully seen in the normalized difference
\begin{equation}
\mathrm{ND}=\frac{w(p_x,p_z)-w(p_x,-p_z)}{w(p_x,p_z)+w(p_x,-p_z)},
\end{equation}  
that would be zero for symmetric spectra (as obtained in dipole approximation). In Fig. \ref{fig7}(b) a high contrast is observed for mid-infrared laser pulses with $\lambda=3400$~nm at a moderate intensity of $I=5\times10^{13}$~W/cm$^2$. This is a clear and easily observable signature of nondipole effects in strong-field ionization. 

\section{Conclusions and Outlook}
We have investigated strong-field photoelectron holography beyond the electric dipole approximation. To obtain reference photoelectron momentum distributions we have solved the TDSE in 2D and 3D including leading order nondipole corrections. In agreement with previous works \cite{chelkowski2015,keil2017} we have found that the holographic interference fingers are shifted in propagation direction of the light: for low momenta $p_x$ in backward and for high momenta $p_x$ in forward direction. The nondipole shift of the fringes becomes more pronounced with increasing laser intensity and wavelength such that for realistic laser mid-infrared light sources the normalized forward-backward difference shows an astonishingly high contrast. The overall $p_x$-dependence of the shift can be explained in a simple man's model based on the three-step model but taking into account the interference between a reference and signal wave. 

In order to calculate the nondipole modifications of the interference pattern in the presence of the long-range Coulomb potential, we have developed a semiclassical model based on the semiclassical approximation of the quantum-mechanical propagator. Compared to other models in dipole approximation \cite{shvetsov2016,lai2015}, an additional term in the semiclassical phase appears that is important for full agreement with the exact  solution of the TDSE. Inspired by the models in Refs. \cite{yan2010,lai2015}, the initial conditions of the classical trajectories and the associated ionization weights are taken from nondipole SFA calculations in saddle-point approximation. Hence, the action of the magnetic field during the tunnel motion \cite{klaiber2013,hartung2019} is included correctly. The resulting model can be used to study quantitatively the interplay between nondipole effects and the Coulomb potential in various laser field geometries, e.g. for different ellipticity or in two-color fields. This paves the way for accurate calculations of PMDs from ionization in mid-infrared fields such as used in Refs. \cite{ludwig2014,maurer2018}.

In contrast to 2D system, where the two-path interference dominates the whole holographic pattern, in real 3D systems, Coulomb focusing plays an important role close to the polarization axis so that a caustic arises in the semiclassical calculation. We have resolved this unphysical structure by extending the glory rescattering approach \cite{xia2018} beyond the dipole approximation. This approach shows that in 3D the (nearly) constructive inference of an infinite number of semiclassical path leads to the pronounced and broad central maximum and it explains the observed shift in propagation direction. 

\begin{acknowledgments}
We thank Nicolas Eicke, Nikolay Shvetsov-Shilovski and Alexander Hartung for valuable discussions. 
\end{acknowledgments}

\bibliography{BrenneckeLein2018a}
\end{document}